\documentclass[12pt,preprint]{aastex}

\def\Mj{\,$\mathrm{M}_\mathrm{J}$}

\begin{document}

\title{HST NICMOS Imaging of the Planetary-mass Companion to 
the Young Brown Dwarf 2MASSWJ\,1207334$-$393254}

\author{Inseok Song\altaffilmark{1}, G. Schneider\altaffilmark{2}, 
        B.~Zuckerman\altaffilmark{3}, J.~Farihi\altaffilmark{1},
        E.~E.~Becklin\altaffilmark{3}, M.~S.~Bessell\altaffilmark{4},
	  P.~Lowrance\altaffilmark{5}, B.~A.~Macintosh\altaffilmark{6}}

\altaffiltext{1}{Gemini Observatory, Northern Operations Center,
                 670 North A'ohoku Place,
                 Hilo, HI 96720, USA, song@gemini.edu,
		     jfarihi@gemini.edu}
\altaffiltext{2}{Steward Observatory, The University of Arizona,
                 933 N. Cherry Ave.,
                 Tucson, AZ 85721 USA, gschneider@as.arizona.edu}
\altaffiltext{3}{Dept. of Physics \& Astronomy and Center for
                 Astrobiology,
                 University of California, Los Angeles,
                 475 Portola Plaza,
                 Los Angeles, CA 90095--1547, USA,
		     ben@astro.ucla.edu, becklin@astro.ucla.edu}
\altaffiltext{4}{Research School of Astronomy and Astrophysics,
                 Institute of Advanced Studies,
                 The Australian National University, ACT 2611,
		     Australia, bessell@mso.anu.edu.au}
\altaffiltext{5}{Spitzer Science Center,
                 Infrared Processing and Analysis Center,
                 MS 220-6, Pasadena, CA 91125, USA,
		     lowrance@ipac.caltech.edu}
\altaffiltext{6}{I Division, Lawrence Livermore National Laboratory,
                 7000 East Avenue,
                 Livermore, CA 94550, bmac@igpp.ucllnl.org}

\begin{abstract}
Multi-band (0.9 to 1.6 $\mu$m) images of the TW\,Hydrae
Association (TWA) brown dwarf, 2MASSWJ\,1207334$-$393254 (also
known as 2M1207), and its candidate planetary mass companion
(2M1207b) were obtained on 2004 Aug 28 and 2005 Apr 26 with
HST/NICMOS. The images from these two epochs unequivocally confirm
the two objects as a common proper motion pair (16.0\,$\sigma$
confidence). A new measurement of the proper motion of 2M1207
implies a distance to the system of $59\pm7$\,pc and a projected
separation of $46\pm5$\,AU. The NICMOS and previously published
VLT photometry of 2M1207b, extending overall from 0.9 to 3.8
$\mu$m, are fully consistent with an object of a few Jupiter
masses at the canonical age of a TWA member ($\sim8$\,Myr) based
on evolutionary models of young giant planets.  These observations
provide information on the physical nature of 2M1207b and
unambiguously establish that the first direct image of a planetary
mass companion in orbit around a self-luminous body, other than
our Sun, has been secured.
\end{abstract}


\section{Introduction}

Beginning in 2004 July, with HST/NICMOS, we initiated a systematic
imaging search for extra-solar gas giant planets around 116 young
nearby stars and brown dwarfs. These targets are $\lesssim50$\,Myr
old and located within 60\,pc of Earth, making them among the best
known targets for such a survey \citep{ARAA}.  

A brown dwarf in the $\sim8$\,Myr old TW~Hydrae Association,
2MASSWJ\,1207334$-$393254 (\citealt{Gizis02}; hereafter 2M1207),
was included in our HST target list and its observation was
planned for 2005 April.  However, on 2004 Apr 27 (UT), 2M1207 was
observed with VLT/NACO and a faint companion candidate was
discovered $\sim0\farcs78$ from the brown dwarf \citep{Chauvin1}.
NICMOS observations of 2M1207 were replanned and brought forward
to 2004 August.  The resulting NICMOS photometric data, shorter in
wavelength than could be obtained with adaptive optics on the VLT,
support the conjecture that 2M1207b is of mid- to late-L type
based upon its color indices.  With the limited precision of the
proper motion data then available for 2M1207 and the short time
between the VLT and 2004 August HST observations, common proper
motion with 2M1207b was established at the $2.6\,\sigma$ level
\citep{Schneider}.  Additional observations were obtained with the
VLT during 2005 February and March that much more precisely
demonstrated common proper motion between 2M1207 and its companion
\citep{Chauvin2}.  As described in Section 4.1, the proper motion
value of 2M1207 \citep{Scholz} used in \cite{Chauvin2} was not
well measured, causing the analysis to be somewhat
over-optimistic. 

With the higher accuracy HST astrometry and a new, more accurate,
proper motion measurement of 2M1207 in the present paper, we
report a more definitive common proper motion between 2M1207 and
2M1207b. We also present short near-IR wavelength diagnostic
photometry which cannot currently be obtained from the ground
given the performance limitations of adaptive optics imaging.

\section{HST Observations}

HST near-IR observations of 2M1207 were obtained with NICMOS
camera 1 ($\sim43$\,mas/pixel) at two observational epochs: 2004
Aug 28 and 2005 Apr 26.  At each epoch, 2M1207 was observed at two
field orientations (spacecraft roll angles) in successive HST
orbits to permit self-subtractions of the rotationally invariant
PSF, thereby significantly increasing the visibility of the nearby
companion.  Direct imaging in camera 1, rather than coronagraphy
in camera 2, was planned due to the $\sim0\farcs78$ angular
separation of 2M1207 and its putative planetary-mass companion and
the relatively benign contrast ratios expected based on the VLT
observations. NICMOS camera 1 provides shorter wavelength
diagnostic filters than camera 2 with a commensurately finer
pixel scale to permit critical sampling of the PSF at short
near-IR wavelengths. Details of the NICMOS observations are listed
in Table~1.

Raw multiaccum frames were converted to count-rate images with an
IDL-based analog to the STSDAS CALNICA task, using calibration
reference files developed by the NICMOS Instrument Definition Team
and the Space Telescope Science Institute.  Known defective pixels
(under-responsive due to particulate contamination, and those with
excessive dark currents) were replaced by two-dimensional Gaussian
weighted interpolation of good neighbor pixels (with wavelength
dependent weighting radii of the PSF FWHM for the filters
employed). Dark subtracted, linearity-corrected, flat-fielded,
cosmic ray rejected count-rate images were post-processed to
remove additional well-characterized detector artifacts. For each
filter in each visit, the four images were astrometrically
registered to the position of the first image.  Position offsets
were taken as reported in the downlinked spacecraft telemetry,
through the CD matrices in the science data file headers, and
verified by Gaussian profile fitting and image centroiding of the
2M1207 PSF image cores.  Image registration was accomplished via
sinc function apodized bicubic interpolation, rebinning the
interpolated images into a two times finer spatially resampled
grid. Then, images were median combined to create better sampled,
higher S/N, and defect minimized count-rate images suitable for
PSF subtraction and of high photometric fidelity. Details of the
data calibration and processing methodologies are discussed in
\citet{Schneider05}.

\subsection{Photometry}

Photometry of 2M1207b was carried out after reducing the light of
the primary star by subtracting the dither combined image at the
second field orientation from the first.  This roll subtraction
virtually eliminated the spatially variable light from the primary
outside of $0\farcs2$, leaving a positive and negative image of
2M1207b displaced by the differential field rotation. Since the
PSFs of the companion partially overlap in this difference image,
model PSFs were created to fit and null out actual image PSFs
separately by adjusting position and flux density. This created an
image at each orientation with the primary removed and only one
companion remaining. These two images of the companion were then
rotated by the appropriate spacecraft roll angle and combined
(Figure 1).  The model PSFs were created using TinyTim 6.1a
\citep{Krist}, which produces high fidelity filter and position
dependent model PSFs for HST instruments.

The in-band flux densities (and Vega system magnitudes) of 2M1207
were established from the unsubtracted, dither combined images
resulting from each orbit independently. To verify the model PSF
subtraction method used to determine the companion magnitudes, we
measured the primary magnitudes both with flux-scaled model PSF
subtraction, and with background subtracted aperture photometry
(corrected to an infinite aperture).

The measured F090M and F160W magnitudes at the two NICMOS
epochs are consistent (Table 2), and flux-weighted mean apparent
magnitudes from both epoch measurments are
$m_{F090M}=22.46\pm0.25$\,mag and $m_{F160W}=18.26\pm0.02$\,mag,
respectively.


\subsection{Astrometry}

The NICMOS camera 1 pixel scales applicable to both epochs of our 2M1207
observations are: X scale = 43.190 mas/pixel, Y scale = 43.016
mas/pixel.  The median combined calibrated count rate images were
geometrically corrected to yield flux-conserved resampled pixels
of 43.190 mas/pixel in both axes. All astrometric measures were
made on the geometrically corrected images.

The location of the primary was measured from the combined
image by Gaussian profile fitting of the primary's PSF core. We
also measured the position of a serendipitously appearing field
star, which could be used for future epoch differential proper
motion measures of 2M1207 itself (Table 3).

The locations of the companion, one from each field orientation at
each epoch, were determined by the position offsets of the model
PSF implants used to null the companion images.  The relative
positions of 2M1207 and its companion were transformed to position
angle (PA; degrees east of north) and angular separation (in mas)
based upon the well established HST/NICMOS focal plane metrology
(aperture orientation in the telescope focal plane), the celestial
orientation of the spacecraft, and the detector X/Y image scales
of NICMOS camera 1.  

\section{2M1207 Color Indices and Implications}

In Figure~\ref{colors}a, $m_{F090M}-m_{F160W}$ and $m_K-m_L$
colors of 2M1207 and b (Table~2) are compared to those of model
calculations -- ``Dusty model'' [\citealt{Dusty}] and ``Clear
model'' [\citealt{Cond}] -- and field M, L, and T dwarfs from
\citet{Leggett}.  Our red colors are consistent with
\cite{Chauvin1} who noted 2M1207b has a very red color
($m_H-m_K=1.16$\,mag) even compared to most known L dwarfs.  Our
colors for 2M1207b fall in between two extreme model cases
indicating that dust clouds significantly affect its atmosphere.
AB~Pic~b, a $\sim13$ Jupiter mass (\Mj) young planetary (or brown dwarf)
companion to the $\sim30$\,Myr old K2V star, AB~Pic
\citep{Chauvin3}, also has unusually red near-IR colors for an L1
dwarf (Figure~\ref{colors}b). A field L-dwarf, 2MASS
J01415823$-$4633574, with very low surface gravity also shows
very red near-IR colors \citep{YoungL}.  In each of these cases,
the unusual redness of these objects must be related to low
surface gravity due to youth and this extreme red color can be
used to identify young brown dwarfs in the solar neighborhood.

With our estimated distance of $59\pm7$\,pc (see Section~5), F090M
\& F160W absolute magnitudes imply a mass of $2-8$\Mj\ for 2M1207b
(``Dusty'' model: $5-8$\Mj\ and ``Clear'' model: $2-5$\Mj). On the
other hand, the best estimated mass of 2M1207b from a color-color
diagram (Figure~\ref{colors}a) is $6-8$\Mj. From these mass
ranges, the mass of 2M1207b is estimated to be $5\pm3$\Mj\ and
the temperature corresponding to this mass range is $920-1540$\,K.
Our estimated mass is consistent with the earlier estimate by
\citet{Chauvin1} from HKL$'$ photometric data ($5\pm2$\Mj)
which was based on a larger distance (70\,pc).  The fact that we
derive the same estimated mass with a smaller distance indicates
that 2M1207b is brighter at shorter wavelengths (e.g., F090M) than
model predictions which implies a bluer $m_{F090M}-m_{F160W}$
color than expected. Unusually red $m_H-m_K$ and blue $m_{F090M}-m_{F160W}$
colors could indicate that 2M1207b is somewhat subluminous in the
$H$ ($F160W$) band compared to models and to field L dwarfs.

\section{Confirmation of Physical Companionship}

\subsection{Improved Proper Motion Determination of 2M1207}

\cite{Scholz} estimated the proper motion\footnote{We note that
two notations are used in the literature for proper motions
in the RA direction; $\mu_\alpha$ and $\mu_\alpha \cos\delta$.
While it is true that the number of arc seconds to go once around
in RA changes as a function of declination (so the need of the
$\cos\delta$ term), that is all that changes. Offsets, proper
motions, etc given in arc second are not dependent on declination.
Therefore, it is not necessary or even correct to use a notation
$\mu_\alpha \cos\delta$ for proper motions given in arc second per
year as in Hipparcos, Tycho-2, UCAC2, etc.  Here we follow
precedent as well established by \citet{Gliese} in his Catalog of
Nearby Stars, and list proper motions in arc seconds per year.} of
2M1207 as, $(\mu_\alpha, \mu_\delta) =(-78\pm11,
-24\pm9)$\,mas/yr, using positions from SuperCOSMOS, 2MASS, and
DENIS catalogs and from the Chandra database.  As there are large
differences in the accuracy between the SuperCOSMOS, 2MASS, and
DENIS positions (e.g., $<60$\,mas for 2MASS versus $<500$\,mas for
SuperCOSMOS), a precise and reliable estimation of proper motion
was difficult to obtain \citep{Mamajek}.  
\citet{Mamajek} estimated proper motions of 2M1207
($\mu_\alpha=-72\pm7, \mu_\delta=-22\pm9$)\,mas/yr using positions
from the same set of catalogs as in \citet{Scholz} excepting the
problematic Chandra pointing position.  Mamajek's calculation
takes into account the positional errors of input catalog
positions and his derived proper motions are overlapping, within
errors, with our proper motion measurements described below.
However, as discussed in Section~5, there are some caveats with
respect to the analysis by \citet{Mamajek}.

A more precise measurement of 2M1207's proper motion was obtained
in the following manner. On 2005 Mar 2, a single 120 second
exposure $I_C$-band image of the field surrounding 2M1207 was
obtained at a parallactic angle of $44.15\arcdeg$ utilizing the
Tektronix $2048\times2048$ CCD camera on the University of Hawaii
2.2 meter Telescope at Mauna Kea Observatory. The measured FWHM of
point sources in the frame was 1$\farcs$1 (5 pixels). An archival
image, observed on 1978 May 2, was retrieved from the SuperCOSMOS
Sky Survey database, where a digitized scan was extracted from a
UK Schmidt Telescope plate.  Because the $I_C$-band frame was
obtained through high airmass (3.3) and 2M1207 is considerably
redder than other objects in the frame, differential atmospheric
refraction could affect the estimated proper motions
\citep{Monet}.  The IRAF task SYNPHOT was used to calculate the
shift in effective wavelength between a typical field star
(average G8) and 2M1207 (M8), resulting in a $175\AA$ shift in the
Cousins $I$-band filter.  At the observed elevation of
$17.7\arcdeg$, this shift from $7890\AA$ to $8065\AA$ yields a
differential refraction of $\approx50$ mas (Howell 2000) which is
small compared to the centroid shift of $\sim1800$\,mas over 27
years. Moreover, Howell's calculation is for an altitude of
2.2\,km whereas our measurement was done at an altitude of 4.1\,km,
thus the true differential atmospheric refraction effect should be
smaller than 50\,mas.

About sixty point sources were selected within the common
$7\farcm5\times 7\farcm5$ field of view of the $I_C$-band image
and the 1978.33 SuperCOSMOS image. These point sources have
centroids that are measured with robust S/N ($\ga10$, unsaturated)
and without obvious proper motion when blinking with the 2005
$I_C$-band image.  The centroid coordinates of the sources
together with those of 2M1207 were entered into the IRAF routine
GEOMAP which calculates a general coordinate transformation,
including rotation and distortion.  The deviation (while excluded
from the calculation) of 2M1207 from the resulting coordinate
transformation should be its relative motion with respect to the
field sources, whose residuals give a measurement of the standard
error.  The resulting proper motion of 2M1207 is calculated to be
$\mu_\alpha=-60.2\pm4.9$\,mas/yr and
$\mu_{\delta}=-25.0\pm4.9$\,mas/yr.

In order to assess the proper motion of 2M1207 without the effects
of differential atmospheric refraction, another I-band image was
obtained on 2006 July 8 (2006.52) at the Siding Spring Observatory
1 meter Telescope using the Wide Field Imager.  A 480 second
exposure was taken at airmass=1.06 (elevation 71 deg) with
measured point source FWHMs=$1\farcs5$ (4 pixels).  At this
elevation, there should be negligible differential atmospheric
refraction.  Using the method described above against the 1978.33
epoch SuperCOSMOS image with 62 point sources in common, we
calculate $\mu_\alpha=-61.0\pm4.6$\,mas/yr and
$\mu_{\delta}=-20.1\pm4.9$\,mas/yr which is in good agreement with
the analysis above.

A final proper motion value is estimated from the weighted average
of the 1978/2005 and 1978/2006 epoch values:
$\mu_\alpha=-60.6\pm3.4$\,mas/yr and
$\mu_{\delta}=-22.6\pm3.5$\,mas/yr. The listed errors at each
epoch are the scatter in field sources produced by GEOMAP.  For a
distance of 59\,pc, the annual parallax of 2M1207 of 17\,mas has
negligible effect on our calculation because the 1978/2005/2006
measurements were obtained during similar seasons of the year.
Furthermore, even for the worst case (17\,mas offset for the 27
year baseline), this effect is much less than the random
centroiding error. 


\subsection{Common proper motion pair}

Measured separations between 2M1207 and 2M1207b from the HST and
VLT are listed in Table~3. Due to the proper motion of the
primary, relative positions between the primary and a stationary
background object should change over time as illustrated in
Figure~3.  Because the 2004 Aug 28 HST observation has the best
positional measurement accuracy, all other positions are plotted
relative to it. If the planetary mass companion candidate seen in
the 2004 Aug 28 image had been a stationary background source,
then its position relative to 2M1207 would have changed along the
black line of Figure~3.  As clearly shown in Figure~3 and Table~3,
2M1207 and its planetary mass companion (2M1207b) are comoving,
indicative of a gravitationally bound pair. 

The weighted mean offset position of 2M1207b relative to 2M1207
from HST and VLT images is $\Delta \alpha=629.8\pm1.3$\,mas and
$\Delta \delta=-448.1\pm1.1$\,mas. Comparing this against the
expected position for a background object at epoch 2005 Apr 26,
$\Delta \alpha=674.6\pm2.2$\,mas, $\Delta
\delta=-415.6\pm2.3$\,mas, we find the two positions differ by
16.2\,$\sigma$. Using only the HST images, the level drops to
16.0\,$\sigma$ which is an unambiguous independent confirmation of
the proper motion companionship of 2M1207 and 2M1207b.
Uncertainties in background object position were calculated using
our improved proper motions and distance ($59$\,pc, see
Section~5 for details on distance estimation).

\section{Discussion}

\citet{Chauvin2} argue that the mass of 2M1207b falls in the
planetary range ($<13.6$\Mj).  To prove this conjecture one must
establish that (1) 2M1207 is very young, i.e., a member of the
$\sim8$\,Myr old TW Hydrae Association and (2) 2M1207 and 2M1207b are
gravitationally bound.

Provided that (1) and (2) are true, then Chauvin et al (2004,
2005) describe how two, essentially independent, techniques --
evolutionary modeling and surface gravity analysis -- both yield
masses for 2M1207b well below 13.6\Mj.  In Figure~3 and Table~3 of
the present paper, we establish with high confidence that 2M1207
and 2M1207b are gravitationally bound.

Thus, only membership of 2M1207 in the TWA remains to be
considered.  Chauvin et al. (2004) list a variety of reasons (their
Section 3.1) why 2M1207 is a member of the TWA.  We can now
strengthen their argument.  On-going Keck NIRSPEC spectroscopic
surveys of brown dwarfs, both old and young, \citep[e.g.,][]
{McGovernThesis, McGovern04, McLean} corroborate earlier
measurements (mentioned in \citealt{Chauvin1}) indicating low
surface gravity in 2M1207, characteristic of 
$\sim10$\,Myr brown dwarf. \citet{Mohanty} report a detection of [O\,I]
emission, which indicates a mass outflow from 2M1207, as well as a
detection of accretion shock-induced UV emission (see their
Section 5.1 for details). All these strongly support that 2M1207
is indeed a young accreting brown dwarf.

Using our new determination of the proper motion of 2M1207, we can
derive a more reliable estimate of the distance to 2M1207 than was
available to Chauvin~et~al. TWA members HR~4796A and 2M1207 lie
very close together in the plane of the sky and, thus, should have
proper motions which differ only according to their relative
distances from Earth. The proper motion of HR~4796A is given in
the Tycho 2 catalog ($\mu_\alpha=-53.3\pm1.3$\,mas/yr,
$\mu_{\delta}=-21.2\pm1.1$\,mas/yr). We used Tycho-2 proper
motions over published, ostensibly more precise, Hipparcos values
because Hipparcos proper motion uncertainties seem to have been
underestimated (e.g., \citealt{Kaplan} and \citealt{Soderblom}).
From the ratio of total proper motions of HR~4796A and 2M1207
(1.144:1.000) and using the Hipparcos measured distance to
HR~4796A ($67.1\pm3.4$\,pc), we deduce a distance to 2M1207 of
$59\pm7$\,pc.  At this distance, the projected separation of
2M1207b from 2M1207 is $46\pm5$\,AU. 

Our proper motion based method to derive the distance to 2M1207
can be tested on TWA 25 which is also close in the plane of the
sky to HR~4796A. The proper motion of TWA~25 is ($-75.9$,
$-26.3$), thus from the ratio of total proper motion to HR~4796A
(1.400:1.000), one derives a distance to TWA~25 of $\sim48$\,pc.
This is in good agreement with the photometric distance of 44\,pc
given in \citet{Song}.  In this case (unlike 2M1207), because
TWA~25 is an M0 dwarf and there are many similar spectral type
members in the TWA, the photometric distance should be fairly
reliable.  Thus, the agreement between the proper motion and
photometrically derived distances for TWA~25, supports the proper
motion method we use to derive the distance to 2M1207.

Based on the moving cluster method, \citet{Mamajek} estimated a
somewhat smaller distance ($53\pm6$\,pc) to 2M1207.  Because the
moving cluster method to the TWA is grounded in good Hipparcos
measured distances to only three stars (TW~Hya, HD~98800, and
HR~4796A; the Hipparcos distance to TWA~9 appears to be wrong),
conclusions drawn in the study of \cite{Mamajek} should be
regarded with care. For example, that study rejects TWA membership
for the nearby ($\sim22$\,pc) mid-M dwarf TWA~22 \citep{Song}
because doing so substantially improves the vertex solution
presented. Yet TWA~22 has a very strong lithium line \citep{Song}
and our survey of nearby, active, M-type stars over most of the
sky (Song, Bessell, \& Zuckerman, in preparation) has revealed
strong lithium lines only in the direction of the TWA with the
exception of a very few M-type members of the $\beta$~Pic moving
group. However, as the $\beta$~Pic moving group has nearly the same
Galactic space motion as the TWA (Table~7, \citealt{ARAA}), even
in the very unlikely event that TWA~22 is a $\beta$~Pic member,
this could hardly impact the vertex solution.  In addition, TWA~22
turns out to be a $\sim0\farcs1$ binary with $\Delta
K\sim0.4$\,mag (Chauvin et~al., in prep.) pushing its photometric
distance from 22\,pc to 28\,pc.  Therefore, TWA~22 is very likely
a true TWA member. Because of such issues, we adopt our
$59\pm7$\,pc distance determination for 2M1207, in preference to
earlier estimations. 

Calculations of opacity-limited fragmentation in a turbulent 3-D
medium yield minimum masses $\gtrsim7$\Mj\ (e.g., \citealt{Low};
\citealt{Boyd}; \citealt{Bate} and references therein). Given the
uncertainties in these calculations and in evolutionary models of
young, planetary-mass objects, our $5\pm3$\Mj\ estimate for
2M1207b is likely not in conflict with the 7\Mj\ fragmentation mass.
Alternatively, perhaps a different model, for example 2-D
fragmentation of a shock compressed layer \citep{Boyd}, will
ultimately be needed to account for the properties of 2M1207b.

\section{Conclusions}

The common proper motion confirmation of the first imaged
planetary mass companion to a celestial object other than our Sun
enables the onset of a new era in extra-solar planet
characterization -- direct spectroscopic analysis.  Absorption
spectroscopy of stellar light reprocessed through atmospheres of
planets detected through radial velocity surveys has been
demonstrated (e.g., HD~209458b; \citealt{Brown}). 2M1207b provides
the first opportunity to collect and spectroscopically analyze
photons from an extra-solar planetary mass companion. Exploiting
the superb stability of the HST, we will attempt to obtain a
near-IR grism spectrum of 2M1207b.  Relative to clear
atmospheric models, dusty models \citep[for example]{Dusty}
predict more flux suppression in the $1.0-1.3 \mu$m range, which
can be readily compared to the anticipated S/N$\sim$10 grism
spectrum.

\acknowledgements{This research was supported by STScI through
grant HST-GO-10176. Portions of this research were performed under
the auspices of the U.S. Department of Energy by the University of
California, Lawrence Livermore National Laboratory under Contract
W-7405-ENG-48.  We thank Michael Wenz (STScI) and Merle Reinhart
(CSC) for their assistance in determining HST's absolute
orientation error following the guide star acquisitions for each
of our orbits. Our gratitude is extended to Al Schultz and Beth
Perrillo (our contact scientist and program coordinator at STScI)
for their assistance in implementing and scheduling our
observations. We thank Richard Crowe for kindly obtaining the 2005
$I$-band data and Bob Shobbrook for the 2006 $I$-band image used
in our proper motion refinement. We thank a referee for pointing
out to us the potential importance of differential atmospheric
refraction in our 2005 $I$-band image.}


\clearpage

\begin{figure}
\begin{center}
\includegraphics[width=0.95\columnwidth]{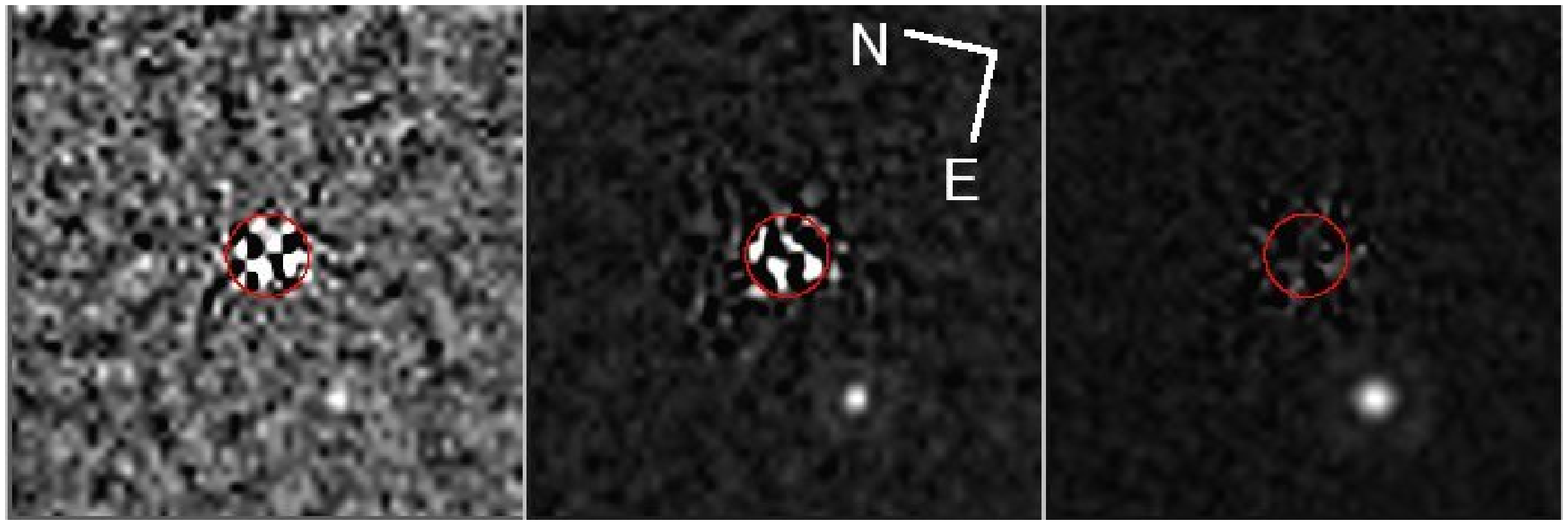}
\end{center}
\caption{
First epoch NICMOS camera 1 PSF-subtracted images of 2M1207
(centered in $0\farcs2$ radius circle) and its companion ($0\farcs77$ to
the SE) at 0.9, 1.1, and 1.6\,$\mu$m (left to right).  By
subtracting a second image of 2M1207 at a different
celestial orientation, background light from the primary at the
location of 2M1207b is effectively eliminated.  Additionally, in
each difference image, a flux-scaled, astrometrically-registered
model PSF, fit to the ``negative'' image of 2M1207b, was added to
eliminate the negative imprint from the difference image.  }
\end{figure}

\clearpage
\begin{figure}
\begin{center}
\includegraphics[width=0.45\paperwidth,keepaspectratio]{f2a.eps} \\
\includegraphics[width=0.45\paperwidth,keepaspectratio]{f2b.eps} \\
\end{center}
\caption{ (a) NICMOS (from this study) and broadband (from
\citealt{Chauvin1}) colors of 2M1207 and 2M1207b are compared to
``Dusty" \citep{Dusty} and ``Clear" \citep{Cond} models. F090M and
F160W magnitudes of field M, L, \& T dwarfs \citep{Leggett} are
calculated using the IRAF task SYNPHOT from their measured
ground-based spectra. (b) Near infrared colors of AB~Pic and b
\citep{Chauvin3} are compared to those of field M, L, \& T dwarfs
\citep{Leggett}. For both 2M1207b (L5 or later) and AB~Pic~b
(L1), their extreme red color is caused by low surface gravity
which is in turn attributed to their young ages. Figure~3 from
\citet{Chauvin1} shows a similar redness of 2M1207b in $HKL'$
colors.
\label{colors}}

\end{figure}

\clearpage
\begin{figure}
\includegraphics[width=0.95\columnwidth]{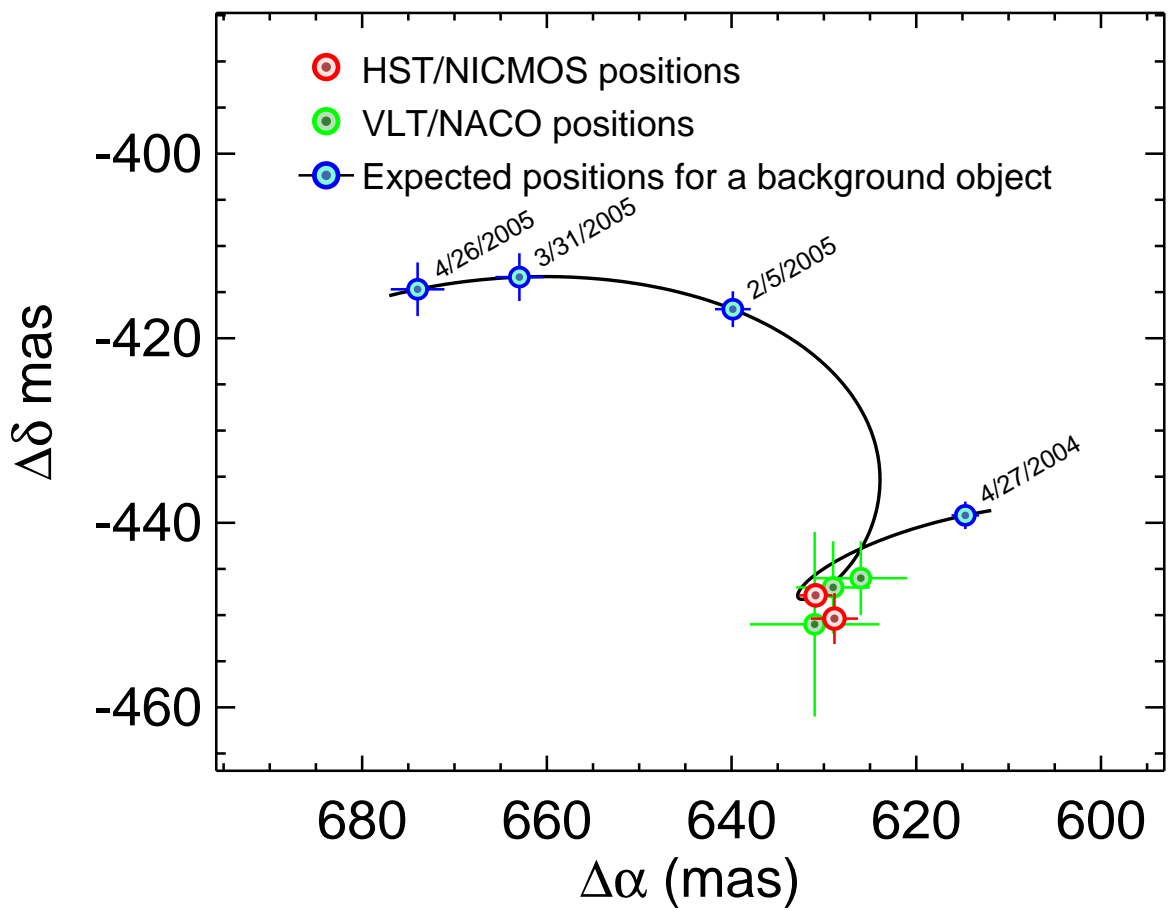}
\caption{Unchanging separation and position angle between 2M1207
and 2M1207b.  Observations with the HST and VLT prove that 2M1207
and its planetary mass companion share common proper motion,
which means they are gravitationally bound. The green and red
crosses mark the observed positions of the companion relative to
2M1207 on different dates. Were the planetary mass candidate seen
in the 2004 Aug 28 image a stationary background object, then its
position, relative to the position measured on 2004 Aug 28, would
have changed as shown by the black line. A distance of 59\,pc was
used to calculate expected background positions.} 
\end{figure}

\clearpage

\def\mc2{\multicolumn{2}{c}}
\def\mc3{\multicolumn{3}{c}}

{\centering
\begin{table}[!H]
\caption{NICMOS Observations Summary}
\begin{tabular}{lccc@{}lccc}
\hline
UT Date     & \mc3{2004 Aug 28}             && \mc3{2005 Apr 26}\\
Orient  $^a$& \mc3{$279\fdg720/289\fdg620$} && \mc3{$187\fdg120/207\fdg120$} \\
\cline{2-4}
\cline{6-8}
Filters      & F090M & F110M & F160W && F090M & F145M & F160W \\
SAMPSEQ  $^b$& STEP64& STEP64& STEP8 && STEP64& STEP64& STEP8 \\
NSAMP    $^c$&   13  &   12  &   12  &&  13   &  12   &  12   \\
EXPTIME  $^d$& 2560s & 2048s &  448s && 2560s & 2048s &  448s \\
\hline
\multicolumn{8}{l}{$a$ Orientation angle of the image +Y axis east of north.}\\
\multicolumn{8}{l}{$b$ Multiaccum sample sequence \citep{Noll}.}\\
\multicolumn{8}{l}{$c$ Number of non-destructive readouts for each exposure.}\\
\multicolumn{8}{l}{$d$ Total exposure time for 4 filtered images at each orientation for each epoch.}\\
\end{tabular}
\end{table}
\par}

\clearpage

\def\mc2{\multicolumn{2}{c}}
\begin{deluxetable}{cccr@{$\pm$}lr@{$\pm$}lr@{$\pm$}l}
\tablecaption{Photometry of 2M1207 and its companion.  }
\tablehead{
 & \mc2{Filter}       & \mc2{HST} & \mc2{HST} \\
 \cline{2-3}
 & Name & 50\% band pass & \mc2{08/28/04} & \mc2{04/26/05} \\
 &      &    ($\mu$m)    & \mc2{(mag)}   & \mc2{(mag)} }
\startdata
2M1207  & F090M & $0.80-1.00$ &  14.66 & 0.03  & 14.71 & 0.04 \\
2M1207  & F110M & $1.00-1.20$ &  13.44 & 0.03  & \mc2{---}    \\
2M1207  & F145M & $1.35-1.55$ &  \mc2{---}     & 13.09 & 0.03 \\
2M1207  & F160W & $1.40-1.80$ &  12.60 & 0.03  & 12.63 & 0.02 \\
2M1207b & F090M & $0.80-1.00$ &  22.34 & 0.35  & 22.58 & 0.35 \\
2M1207b & F110M & $1.00-1.20$ &  20.61 & 0.15  & \mc2{---}    \\
2M1207b & F145M & $1.35-1.55$ & \mc2{---}      & 19.05 & 0.03 \\
2M1207b & F160W & $1.40-1.80$ &  18.24 & 0.02  & 18.27 & 0.02 \\
\enddata

\tablecomments{\small The uncertainties in the NICMOS magnitudes include
error estimates from the model PSF fitting (companion) and
aperture photometry (primary) as well as the uncertainties in the
absolute photometric calibration of the instrument in each filter
band for camera 1. \citet{Chauvin1} reported JHKL$'$ photometry of
2M1207 (J$=13.00\pm0.03$\,mag, H$=12.39\pm0.03$\,mag,
K$=11.95\pm0.03$\,mag, \& L$'=11.38\pm0.10$\,mag) and 2M1207b
(H$=18.09\pm0.21$\,mag, K$=16.93\pm0.11$\,mag, and
L$'=15.28\pm0.14$\,mag).
The NICMOS F160W filter is $\sim30$\% wider than the ground-based
Johnson H-band filters, so comparison of our data with previously
published $H$-band data requires a careful conversion (see
\citealt{Patrick}, for details).} 
\end{deluxetable}

\def\mc2{\multicolumn{2}{c}}
\begin{deluxetable}{ccr@{$\pm$}lr@{$\pm$}ll}
\tablecaption{Differential astrometry of 2M1207b relative to 2M1207.}
\tablehead{
&  Epoch  &\mc2{Separation} &\mc2{Position Angle} & Instrument/Camera\\
&  (UT)   &\mc2{(mas)}   &\mc2{(deg E of N)}  & }
\startdata
2M1207b & 28 Aug 2004 &  773.7 & 2.2  & 125.37 & 0.03 & HST/NICMOS\\
2M1207b & 26 Apr 2005 &  773.5 & 2.3  & 125.61 & 0.20 & HST/NICMOS\\
2M1207b & 27 Apr 2004 &  772   & 4    & 125.4  & 0.3  & VLT/NACO\\
2M1207b & 05 Feb 2005 &  768   & 5    & 125.4  & 0.3  & VLT/NACO\\
2M1207b & 31 Mar 2005 &  776   & 8    & 125.5  & 0.3  & VLT/NACO\\ 
\cline{1-6}
\mc2{Weighted mean (HST + VLT)} &  773.0 & 1.4  & 125.37 & 0.03 & \\ \\
Field Star & 28 Aug 2004 & 7795.3 & 3.1  & 125.33 & 0.01  &\\
Field Star & 26 Apr 2005 & 7848.9 & 3.4  & 125.08 & 0.06  &\\
\enddata
\tablecomments{
The uncertainties in the HST position measurements include known
calibration errors in the NICMOS science instrument aperture
frame, image centroiding, and the absolute error in the spacecraft
orientation as determined from ancillary engineering telemetry
downlinked from the HST pointing control system.  Listed positions
are for the F160W image centroids which are favored over other
bands because; (a) they were identically observed in this highest
S/N bandpass at both epochs, and (b) F160W PSFs
($\lambda/D\sim0\farcs14$) are better sampled with the camera 1
pixel scale (43 mas/pixel) than the shorter wavelength
observations.  Positions were measured in all filter bands, and
while all are consistent within their errors, the
others are not quite as well determined as those made with F160W.}
\end{deluxetable}

\end{document}